%% file: Paper-0009.tex
\documentclass[runningheads]{llncs}
\usepackage[T1]{fontenc}
\usepackage{graphicx}
\usepackage{subcaption}
\usepackage{multirow}
\usepackage[raggedrightboxes]{ragged2e}
\usepackage{float}
\usepackage{hyperref}
\usepackage{amsmath}

\usepackage{xcolor}
\hypersetup{
    colorlinks,
    linkcolor={blue!50!black},
    citecolor={blue!50!black},
    urlcolor={blue!50!black}
}

\usepackage{colortbl} 
\newcolumntype{P}[1]{>{\centering\arraybackslash}p{#1}}
\usepackage{color}

\begin{document}
\title{Quantifying the Impact of Population Shift Across Age and Sex for Abdominal Organ Segmentation}

\titlerunning{Impact of Population Shift for Abdominal Segmentation}

\author{Kate \v{C}evora\inst{1} \and
Ben Glocker\inst{1}\and
Wenjia Bai\inst{1,2,3}}

\authorrunning{K. \v{C}evora et al.}

\institute{Department of Computing, Imperial College London, UK\\ \and
Department of Brain Sciences, Imperial College London, UK \and
Data Science Institute, Imperial College London, UK
\email{\{kc2322,b.glocker,w.bai\}@imperial.ac.uk}
}
\maketitle              
\begin{abstract}
Deep learning-based medical image segmentation has seen tremendous progress over the last decade, but there is still relatively little transfer into clinical practice. One of the main barriers is the challenge of domain generalisation, which requires segmentation models to maintain high performance across a wide distribution of image data. This challenge is amplified by the many factors that contribute to the diverse appearance of medical images, such as acquisition conditions and patient characteristics. The impact of shifting patient characteristics such as age and sex on segmentation performance remains relatively under-studied, especially for abdominal organs, despite that this is crucial for ensuring the fairness of the segmentation model. We perform the first study to determine the impact of population shift with respect to age and sex on abdominal CT image segmentation, by leveraging two large public datasets, and introduce a novel metric to quantify the impact. We find that population shift is a challenge similar in magnitude to cross-dataset shift for abdominal organ segmentation, and that the effect is asymmetric and dataset-dependent. We conclude that dataset diversity in terms of known patient characteristics is not necessarily equivalent to dataset diversity in terms of image features. This implies that simple population matching to ensure good generalisation and fairness may be insufficient, and we recommend that  fairness research should be directed towards better understanding and quantifying medical image dataset diversity in terms of performance-relevant characteristics such as organ morphology.

\end{abstract}

\section{Introduction}
Automated medical image segmentation models have seen tremendous progress in terms of segmentation speed and accuracy, in some cases surpassing the performance of human experts \cite{hatamizadeh2022unetr,Isensee2021,milletari2016v}. However, there is a large gap at present between the plethora of automated segmentation models which are developed in research environments, and those which are integrated into clinical practice. A commonly cited reason for this gap is the often poor generalisation performance of segmentation models to test data which is outside of the distribution of the training data, known as domain shift \cite{zhou2022domain}.

\begin{figure}
\centering
\includegraphics[width=0.8\textwidth]{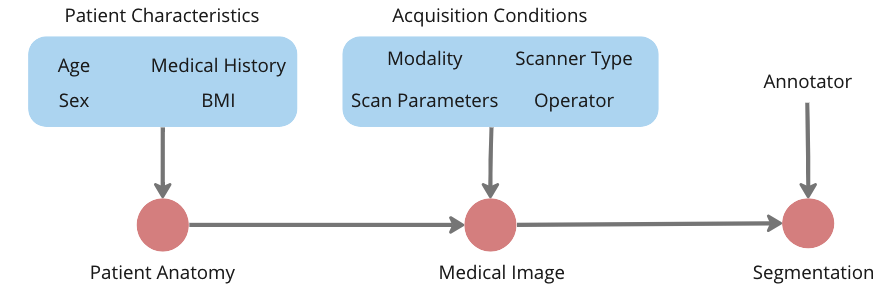}
\caption{Causal diagram illustrating major factors that can influence medical image appearance and associated segmentation. The factors can be split into three broad groups: patient characteristics which directly influence patient anatomy, acquisition conditions which influence image appearance, and annotation protocol which influences manual segmentation style.} \label{causal_diagram}
\end{figure}

When we look at domain shift in medical image segmentation via the lens of causality \cite{castro2020causality}, three common types of shift exist, namely population shift, acquisition shift and annotation shift, illustrated by the casual graph in Figure \ref{causal_diagram}. Population shift is caused by changes in the distribution of patient characteristics such as age, sex, ethnicity and disease prevalence  \cite{yang2023change}. It is particularly important because it has the potential to result in biased model predictions for different patient populations. While acquisition and annotation shift have received significant attention leading to a range of advanced augmentation approaches, domain adaptation methods \cite{Chen2022} and standard operating procedures for annotation \cite{Petersen2016} to mitigate their effects, population shift receives relatively less research attention, in particular for abdominal organ segmentation.

To better understand the influence of population shift on abdominal organ segmentation, we collate a large-scale abdominal CT dataset of 1,582 subjects from public sources along with their population characteristics. We perform the first study to evaluate the impact of population shift with respect to age and sex on segmentation performance for major abdominal organs, and introduce a novel metric, the performance gap, to quantify the maximal impact of population shift for each subgroup of interest. We also compare the impact of population shift on segmentation performance to that caused by cross-dataset shift. Furthermore, we propose a novel hypothesis that the segmentation performance is more directly determined by the training set diversity in terms of image features, rather than population characteristics. We believe that our findings, the evaluation framework and our recommendations for the direction of future research will provide useful insights for the community to elucidate the complex causes and magnitude of population shift in medical image segmentation problems.

\section{Background and Related Works}

\subsubsection{Domain Shift} is a significant challenge for medical image segmentation, occurring when there is a significant shift in the statistical distribution of the appearance of medical imaging data across different sources. Figure \ref{causal_diagram} shows a simplified causal perspective on the factors contributing to image appearance and corresponding segmentation, which can be broadly split into three groups: patient characteristics, acquisition conditions and annotation style. Changes in these factors manifest in medical images in the form of differing anatomical shapes, contrasts, intensity distribution, resolution, or noise patterns. As a result, segmentation models trained on one dataset may not generalise well across data from different sources \cite{hesamian2019deep,isensee2019nnu,ma2021cutting}.

\subsubsection{Population Shift} is a specific type of domain shift which is caused by changes in the relative proportion of subgroups in a dataset \cite{koh2021wilds}. In the context of medical image datasets, subgroups are generally defined by patient characteristics such as age, sex, ethnicity or medical history. Several recent works demonstrate bias in image classification models arising from population shift with respect to sex and ethnicity \cite{larrazabal2020gender,gichoya2022ai,wang2023bias}. This is particularly concerning because under-performance on certain populations at test-time can potentially lead to worse health outcomes for these groups.

There are relatively fewer works examining the impact of population shift on segmentation performance. Ioannou et al. \cite{ioannou2022study} find significant race and sex bias with respect to accuracy for segmentation models trained on unbalanced brain imaging datasets. Lee et al. \cite{lee2022systematic} found that segmentation models trained on cardiac MR images performed worse on racial groups which were underrepresented in the training data.

\subsubsection{Remaining Challenges: }
Despite evidence that population shift can have a significant impact on the performance of medical image segmentation models \cite{ioannou2022study,lee2022systematic}, it is relatively under-studied compared to the impact of acquisition shift. For example, we are unaware of any other works that investigate the impact of population shift with respect to age and sex on segmentation of abdominal organs. Further, for organs and modalities where this impact has been quantified \cite{ioannou2022study,lee2022systematic}, the underlying causal mechanism of this bias has not been investigated. Gaining an understanding of the mechanisms of how population shift leads to change of performance is crucial for designing methods, such as data augmentation strategies, to mitigate its potential impact.

\section{Method}
\subsection{Data}
Although numerous efforts have been devoted to curating large-scale abdominal CT datasets \cite{Li2023well}, most of them do not release patient characteristics. After communicating with the owners of 13 public abdominal CT datasets, we were able to obtain patient-level demographic information for three. Two of them, TotalSegmentator (TS) \cite{wasserthal2022totalsegmentator} and AMOS \cite{ji2022amos}, were sufficiently large to allow sex- and age-based resampling of training datasets to investigate the impact of population shift, which we will use for this work. Further details about the datasets are included in the Supple. Table 1, and will be released with the paper.

\subsection{Experimental Design}
We investigate the effects of population shift with respect to sex and age on segmentation performance for four abdominal organs: the left and right kidneys, pancreas and liver. Changes in shape and composition of these organs with respect to sex and age are known to occur \cite{chouker2004estimation,gava2011gender,kipp2019normal,kreel1973changes,marcos2015liver,sabolic2007gender,wasserthal2022totalsegmentator,zhou2023multi}, making them interesting candidates for investigation. Additionally, we perform a cross-dataset shift experiment to understand the magnitude of population shift in comparison to cross-dataset shift, the latter being significantly better-studied in the domain generalisation literature \cite{guan2021domain,zhou2022domain}.

\subsubsection{Measuring the Impact of Population Shift:}
We construct two subgroups, $g_1$ and $g_2$, for each patient characteristic (sex or age) by sampling without replacement from the full dataset (TotalSegmentator or AMOS). For sex, one subgroup contains only male subjects and the other contains only female. For age, one subgroup contains only subjects under 50 years old and the other contains only subjects over 70 years old. Each subgroup is further split into training and test sets. We train a segmentation model using the training set from a single subgroup, and then evaluate the trained model on the test sets from both subgroups.

To quantify the impact of population shift, we propose a new metric, the \textit{performance gap} $\Delta P$, which measures the change of segmentation performance, e.g. Dice score or 95-percentile Hausdorff distance (HD95), caused by the maximal shift of training set characteristics. The performance gap is normalised by the average segmentation performance and formulated as,
\begin{equation} \label{eq:1}
    \Delta P_{g_1}(g_1, g_2) = \frac{P(g_1, S(g_2)) - P(g_1, S(g_1))}{0.5 \times [ P(g_1, S(g_1)) + P(g_1, S(g_2))]} \times 100\%
\end{equation}
where $P(g_1, S(g_1))$ denotes the performance of a segmentation model $S$ trained on subgroup $g_1$ and tested on subgroup $g_1$, $P(g_1, S(g_2))$ denotes the performance of a model trained on subgroup $g_2$ and tested on subgroup $g_1$, and $\Delta P_{g_1}$ denotes their performance gap when deployed on subgroup $g_1$. Similarly, we can define the performance gap $\Delta P_{g_2}$ when deployed on subgroup $g_2$.

The significance of a performance gap is calculated as a t-test carried out between $P(g_1, S(g_1))$ and $P(g_1, S(g_2))$.

\subsubsection{Measuring the Impact of Cross-Dataset Shift:}
To understand the magnitude of population shift compared to other major sources of domain shift, we investigate the impact of cross-dataset shift. We construct two subgroups sampled from the TotalSegmentator \cite{wasserthal2022totalsegmentator} and AMOS \cite{ji2022amos} datasets respectively. We control for sex and age so that the two subgroups have similar population distributions, meaning that the remaining sources of shift between the two subgroups are mainly scanner, site, study type and disease type. We train segmentation models and assess the performance gap under cross-dataset shift using the same definition Eq.~\eqref{eq:1}, where $g_1$ is formed of subjects from AMOS, and $g_2$ is formed of subjects from TotalSegmentator.

\subsubsection{Measuring Training Data Diversity: } To measure the diversity of the training data, we define a proxy measure of diversity, using the standard deviation of the organ volumes calculated across the training subjects in each subgroup.

\subsubsection{Implementation Detail:}
We use a state-of-the-art 3D nnU-Net \cite{Isensee2021} as the segmentation model, although other architectures can also be used. nnU-Net appears regularly in the leaderboards of recent medical image segmentation challenges \cite{antonelli2022medical,ji2022amos}, and it has an established image pre-processing and data augmentation pipeline. For fair comparison, we ensure that the training set size is the same for both subgroups of a given dataset. Training set sizes for each experiment can be found in the Supple. Table 4. The validation set for parameter tuning is automatically selected by nnU-Net from the training samples. We employ 5-fold cross validation with a hold-out test set for each experiment and report the average results across the folds.

\section{Results}

\begin{table}[h]
    \centering
    \begin{tabular}{|P{1.7cm}|P{2.5cm}|P{1.9cm}|P{1.9cm}||P{1.9cm}|P{1.9cm}|}
        \hline
        \multicolumn{2}{|c|}{} & \multicolumn{2}{|c||}{\textbf{$\Delta P$, Dice (\%)}} & \multicolumn{2}{|c|}{\textbf{$\Delta P$, HD95 (\%)}} \\
        \hline
        \textbf{Dataset} & \textbf{Organ} & \textbf{$g_1$=Female} & \textbf{$g_2$=Male}  & \textbf{$g_1$=Female} & \textbf{$g_2$=Male} \\
        \hline
        & R. kidney & 3.57  & \cellcolor{red!25}-5.94 & 20.5  & 37.1  \\
        TS & L. kidney& 2.45  & \cellcolor{red!25}-6.17 & -10.6  & \cellcolor{red!25}95.3 \\
        & Liver & 1.61  & -0.67 & 21.7  & 23.1 \\
        & Pancreas & 4.15  & -2.79  & -10.4  & 11.2  \\
        \hline
        & R. kidney & 0.27  & -0.11 & -14.3  & -2.13  \\
        AMOS & L. kidney& \cellcolor{green!25}1.25 & -0.42 & \cellcolor{green!25}-119.3 & 89.8  \\
        & Liver & -2.63  & -0.23 & -8.7  & -30.7  \\
        & Pancreas & 1.40  & -1.64 & -22.3  & -7.5  \\
        \hline
        \hline
        
        \textbf{Dataset} & \textbf{Organ} & \textbf{$g_1$=U50} & \textbf{$g_2$=O70}  & \textbf{$g_1$=U50} & \textbf{$g_2$=O70} \\
        \hline
        & R. kidney & -0.38  & 0.19 & 62.4  & 89.0  \\
        TS & L. kidney& 1.65  & -1.67 & \cellcolor{green!25}-124.7 & 108.8  \\
        & Liver & -0.87  & 0.18  & 12.4  & 4.8  \\
        & Pancreas & 1.11  & 3.10  &  -10.2  & 29.1  \\
        \hline
        & R. kidney & 0.48  & -0.23 & -42.4  & -52.9  \\
        AMOS & L. kidney& \cellcolor{green!25}1.04 & -0.25 & -132.7  & -5.2  \\
        & Liver & -0.72  & -0.67  & 41.3  & -1.1  \\
        & Pancreas & 0.44  & -1.99 & -0.6  & 4.3  \\
        \hline
        \hline
        
        \textbf{Dataset} & \textbf{Organ} & \textbf{$g_1$=AMOS} & \textbf{$g_2$=TS}  & \textbf{$g_1$=AMOS} & \textbf{$g_2$=TS} \\
        \hline
        & R. kidney  & 0.46  & -1.12 & 16.7  & \cellcolor{red!25}134.1 \\
        TS/AMOS & L. kidney  & -3.57  & -3.90 & \cellcolor{red!25}135.2 & \cellcolor{red!25}112.1\\
        & Liver & 0.41  & \cellcolor{red!25}-4.66 &  66.3  & \cellcolor{red!25}151.4  \\
        & Pancreas  & 0.24  & \cellcolor{red!25}-10.7 &  29.7  & \cellcolor{red!25}101.3\\
        \hline
    \end{tabular}
    \caption{Performance gaps $\Delta P$ in terms of Dice score and 95 percentile Hausdorff distance (HD95) due to population shift and cross-dataset shift. Coloured cells indicate that the performance gap is statistically significant ($p < 0.05$) via a t-test (N = group size, see Supple. Table 4), with red indicating a negative performance gap and green indicating a positive performance gap.
    Note that for Dice, a negative value indicates worse performance when the training set does not match the test set and for HD95, this is indicated by a positive value. TS: TotalSegmentator; U50: under 50; O70: over 70.}
    \label{tab:main_results}
\end{table}

\begin{figure}[h]
    \centering
    \includegraphics[width=\textwidth]{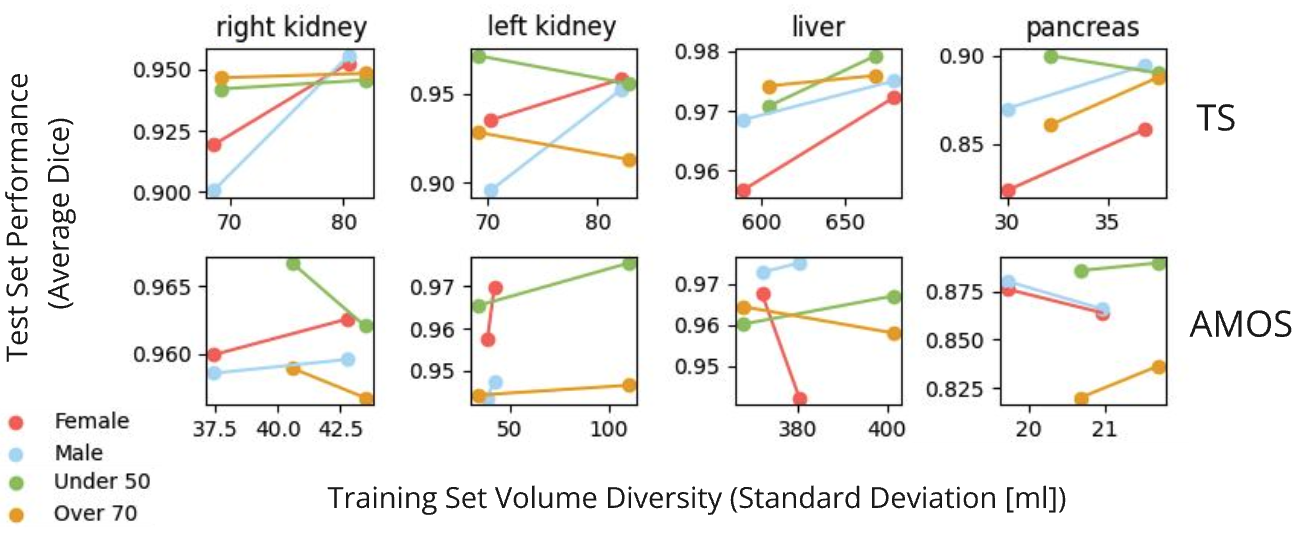}
    \caption{Plots of segmentation performance in terms of Dice score on the test set against the proxy measure of training set diversity, the standard deviation of organ volumes. The test set data has been split by colour-coded subgroups. The top row reports results on the TotalSegmentator dataset (TS) and the bottom row reports results on AMOS.}
    \label{fig:diversity_and_dice}
\end{figure}

Table \ref{tab:main_results} reports the observed performance gaps per dataset, organ and subgroup, along with the results for cross-dataset shift. A green fill indicates significant better performance when the test set matches the training set (positive value for Dice, negative value for HD95) and a red fill indicates significant worse performance when the test set matches the training set (negative value for Dice, positive value for HD95). Figure \ref{fig:diversity_and_dice} shows the test set performance in terms of average Dice plotted against the organ volume diversity of the training dataset, split by subgroup. Raw average Dice scores for each experiment can be found in Supple. Table 3. Below we summarise the main findings:

\subsubsection{The impact of population shift is significant for kidney segmentation.} We see significant performance gaps in terms of both Dice and HD95 for the kidneys under population shift with respect to age and sex. This gap is particularly large for the male kidneys, where we see a performance drop of around 6\% for Dice, and 95\% for HD95. The magnitude of the significant performance gaps across organs observed for population shift (1-6\% Dice, 95-125\% HD95) is similar to that observed for cross-dataset shift (5-11\% Dice, 100-135\% HD95). 

\subsubsection{Proportionate representation of subgroups defined by age and sex is not sufficient to ensure the best performance for these groups at test-time.} Our results show that in some cases, a complete lack of representation in the training data can surprisingly result in better test-time performance compared to when the training and test data match in terms of population characteristics. For the female subgroup sampled from the AMOS dataset, test time performance on the left kidney is significantly better in terms of Dice (1.25\%) and HD95 (-119\%) when the training dataset is fully male, compared to when it is fully female. This is also true for the left kidney in the under 50 subgroup for both AMOS and TotalSegmentator (TS).

\subsubsection{Proxy measure of training data diversity may positively correlate with segmentation performance.} Figure \ref{fig:diversity_and_dice} shows that increased diversity in the training dataset in terms of organ volume standard deviation correlates with increased test-set segmentation performance, in particular on the TotalSegmentator dataset, and possibly for the left and right kidneys on the AMOS dataset. Statistics per subgroup for training the set can be found in Supp. Tab. 2.

\subsubsection{The performance gap is asymmetric between subgroups.} In cases where we see a significant performance gap for one subgroup (e.g TS male kidneys), the complementary subgroup (TS female kidneys) does not necessarily show a similar performance gap. The TS male kidneys have a larger standard deviation of volumes compared with the female subgroup (80mL compared to 69mL), indicating greater diversity, which may explain this asymmetric performance gap. This asymmetry can also be observed in the cross-dataset shift experiments, where training a model with AMOS images causes a significant drop in performance for TS test images, but the same is not true for the AMOS test images. 

\section{Discussion and Future Directions}
This is the first study quantifying the potential impact of population shift with respect to age and sex on the performance of abdominal CT image segmentation, using a state-of-the-art models with a standard set of image augmentations. Our results demonstrate that the impact of population shift with respect to age and sex is significant and can be comparable in magnitude to that caused by cross-dataset shift. This implies that the standard image augmentations employed by many image pre-processing pipelines such as rotation, scaling and random deformation, are insufficient to mitigate these effects. In order to simulate truly diverse abdominal CT datasets, we likely need more advanced image augmentation methods which can simulate real morphological differences between subgroups.

A common and well-supported hypothesis is that under-representation of a subgroup in training data can lead to decreased performance at test-time \cite{ioannou2022study,larrazabal2020gender,lee2022systematic}. However, we have observed that for female kidneys, test-time segmentation performance could be improved by using a male training dataset. These findings are important because not only do they demonstrate potential for bias against certain groups under population shift, they imply that population-matching between training and test data is not sufficient to ensure fairness.

We hypothesise that this outcome may be the result of an imperfect correlation between diversity in terms of patient labels (e.g. sex, age, ethnicity) and diversity in terms of raw image features such as organ morphology, volume and texture. For example, we have observed in this case the male training images showed greater diversity in terms of organ volume than the female dataset, which may explain why the male-trained segmentation model showed better generalisation ability, even outperforming a female-trained model on female images.

We conclude that the impact of population shift with respect to age and sex on performance is significant for abdominal CT segmentation. Proportionate representation of subgroups defined by age and sex is not sufficient to ensure equitable performance at test-time. An initial look at the correlation between training dataset diversity in terms of organ volumes and segmentation performance suggests that measurements of diversity derived from raw image features are likely an important indicator of generalisation performance across subgroups.

In terms of future directions, our findings call for the development of methods to understand and measure medical image dataset diversity directly from raw image-level features such as shape, texture and volume. Such a metric will allow us to build training datasets and design image augmentation methods for medical image segmentations that result in better generalisation across a range of subgroups, without requiring per-patient demographic information. It will also enable predictions of whether a particular dataset is likely to result in a trained segmentation model that shows good test-time generalisation.

\subsubsection{Limitations:}
We have attempted to control the effect of other potentially confounding variables (such as acquisition site, scanner type and study type) on our results by matching distributions of these variables as closely as possible between paired subgroups. However, successfully studying the effect of just a single variable in isolation on segmentation performance is a near-impossible task. Whilst it is theoretically possible to control for some known potentially confounding variables when designing experiments, many more are unknown or unreported. This aligns with our recommendation that fairness research in medical image analysis should be directed at better understanding and improving diversity in terms of performance-relevant characteristics, circumventing the need for detailed patient-level labels.

\begin{credits}
\subsubsection{\ackname} This project was part-funded by the EPSRC CDT in Medical Imaging at King's College London and Imperial College London (EP/S022104/1).

\subsubsection{\discintname}
The authors have no competing interests to declare that are
relevant to the content of this article.
\end{credits}

\bibliographystyle{splncs04}
\bibliography{Paper-0009}

\include{supplementary}

\end{document}

%% file: supplementary.tex
\section*{Supplementary Material}

\begin{table}[H]
\caption{Comparison of key attributes from the two training datasets, TotalSegmentator and AMOS. }\label{dataset_info}
\begin{tabular}{|P{3cm}|P{4.5cm}|P{4.5cm}|}
\hline
 & \textbf{AMOS} & \textbf{TotalSegmentator} \\
\hline
\textbf{Modality} & CT & CT \\
\hline
\textbf{Subjects} & 500 & 1082 \\
\hline
\textbf{Disease Types} & Abdominal tumors/abnormalities & Mixture \\
\hline
\textbf{No.o. Labelled Structures} & 15 & 104 \\
\hline
\textbf{Number of Scanners} & 5 & 16 \\
\hline
\textbf{Number of Sites} & 2 & 9 \\
\hline

\textbf{Slice Spacing} & 1.25-5.0mm (non-isotropic) &  1.5mm (isotropic) \\
\hline
\textbf{Per-patient Attributes} & Sex, Age, Scanner Model, Scanner Manufacturer, Acquisition Date, Site
& Sex, Age, Institute \\
\hline
\end{tabular}
\end{table}

\begin{table}[H]
    \caption{Standard deviation of the training set organ volumes per group ($g_1$ or $g_2$). The grey cells indicate which subgroup has larger variation in the organ volumes. U50=under 50, O70=over 70, TS=TotalSegmentator.}
    \begin{subtable}[h]{\textwidth}
    \centering
    \begin{tabular}{|P{1.7cm}|P{2.5cm}|P{3cm}|P{3cm}|}
        \hline
        \multicolumn{2}{|c|}{} & \multicolumn{2}{|c|}{\textbf{Volume Standard Deviation (ml)}} \\
        \hline
        \textbf{Dataset} & \textbf{Organ} &\textbf{$g_1$=Female} & \textbf{$g_2$=Male}  \\
        \hline
         & left kidney & 69 & \cellcolor{gray!15}80\\
         TS & right kidney & 70 & \cellcolor{gray!15}82 \\
         & liver & 589 & \cellcolor{gray!15}680 \\
         & pancreas & 30 & \cellcolor{gray!15}37 \\
         \hline
         & left kidney & 37 & \cellcolor{gray!15}43 \\
         AMOS & right kidney & 39 & \cellcolor{gray!15}42\\
         & liver & 372 & \cellcolor{gray!15}380\\
         & pancreas & \cellcolor{gray!15}21 & 20 \\
         \hline
         \hline
        \textbf{Dataset} & \textbf{Organ} & \textbf{$g_1$=U50} & \textbf{$g_2$=O70}  \\
        \hline
         & left kidney & \cellcolor{gray!15}82 & 69\\
         TS & right kidney & \cellcolor{gray!15}83 & 69\\
         & liver & \cellcolor{gray!15}669 & 604\\
         & pancreas & \cellcolor{gray!15} 38 & 32 \\
         \hline
         & left kidney &\cellcolor{gray!15} 43 & 41\\
         AMOS & right kidney & 34 & \cellcolor{gray!15}111\\
         & liver & \cellcolor{gray!15}402 & 367\\
         & pancreas & 21 & \cellcolor{gray!15}22 \\
         \hline      
    \end{tabular}
    \end{subtable}
    \label{tab:volume_results}
\end{table}
    \pagebreak

\begin{table}
\begin{subtable}[h]{\textwidth}
\ContinuedFloat
    \centering
    \begin{tabular}{|P{1.7cm}|P{2.5cm}|P{3cm}|P{3cm}|}         
         \hline
        \textbf{Dataset} & \textbf{Organ} & \textbf{$g_1$=AMOS} & \textbf{$g_2$=TS} \\
        \hline
         & left kidney & 51 & \cellcolor{gray!15}83  \\
         TS/AMOS & right kidney & \cellcolor{gray!15}109 & 85 \\
         & liver & 423 & \cellcolor{gray!15}645 \\
         & pancreas & 22 & \cellcolor{gray!15}37  \\
         \hline
    \end{tabular}
    \end{subtable}
    \label{tab:volume_results}
\end{table}

\setcounter{table}{2}

\begin{table}[H]
    \caption{Average Dice score per organ (with standard deviation in brackets) for each test set group (Ts=$g1$ or Ts$g2$) using a model that was trained on either $g1$ (Tr=$g_1$) or $g2$ (Tr=$g_2$).}
    \label{tab:main_results}
    \centering
    \begin{tabular}{|p{1.7cm}|p{2.5cm}|p{1.9cm}|p{1.9cm}||p{1.9cm}|p{1.9cm}|}
        \hline
        \multicolumn{2}{|c|}{} & \multicolumn{2}{|c||}{\textbf{Ts = $g_1$}} & \multicolumn{2}{|c|}{\textbf{Ts = $g_2$}} \\
        \hline
        \textbf{Dataset/ Group} & \textbf{Organ} & \textbf{Tr=$g_1$} & \textbf{Tr=$g_2$}  & \textbf{Tr=$g_2$} & \textbf{Tr=$g_1$} \\
        \hline
        TS              & right kidney  & 0.93 (0.21) & 0.96 (0.20)& 0.95 (0.24) & 0.89 (0.22) \\
        $g_1$=Female    & left kidney   & 0.92 (0.19) & 0.95 (0.16)& 0.96 (0.23) & 0.90 (0.23) \\
        $g_2$=Male      & liver         & 0.96 (0.09) & 0.97 (0.13)& 0.97 (0.05) & 0.97 (0.05) \\
                        & pancreas      & 0.82 (0.23) & 0.86 (0.21)& 0.90 (0.17) & 0.87 (0.20) \\
        \hline
        
        AMOS            & right kidney  & 0.96 (0.04) & 0.96 (0.03)& 0.96 (0.03) & 0.96 (0.04) \\
        $g_1$=Female    & left kidney   & 0.96 (0.03) & 0.97 (0.05)& 0.94 (0.12) & 0.95 (0.12) \\
        $g_2$=Male      & liver         & 0.97 (0.03) & 0.94 (0.04)& 0.97 (0.02) & 0.97 (0.03) \\
                        & pancreas      & 0.86 (0.07) & 0.88 (0.09)& 0.87 (0.10) & 0.86 (0.09) \\

        \hline
        \hline
        
        TS          & right kidney      & 0.95 (0.15) & 0.94 (0.16)& 0.95 (0.16) & 0.95 (0.13) \\
        $g_1$=U50   & left kidney       & 0.96 (0.14) & 0.97 (0.09)& 0.93 (0.21) & 0.91 (0.22) \\
        $g_2$=O70   & liver             & 0.98 (0.04) & 0.97 (0.07)& 0.97 (0.06) & 0.98 (0.05) \\
                    & pancreas          & 0.89 (0.18) & 0.90 (0.16)& 0.86 (0.23) & 0.89 (0.17) \\

        \hline
        AMOS        & right kidney      & 0.97 (0.02) & 0.96 (0.02)& 0.96 (0.03) & 0.96 (0.03) \\
        $g_1$=U50   & left kidney       & 0.97 (0.02) & 0.97 (0.02)& 0.94 (0.06) & 0.94 (0.10) \\
        $g_2$=O70   & liver             & 0.96 (0.06) & 0.97 (0.04)& 0.96 (0.08) & 0.96 (0.06) \\
                    & pancreas          & 0.89 (0.06) & 0.89 (0.07)& 0.82 (0.21) & 0.84 (0.17) \\

        \hline
        \hline
         & right kidney & 0.96 (0.11) & 0.96 (0.06) & 0.92 (0.21) & 0.91 (0.17) \\
         $g_1$=TS & left kidney & 0.96 (0.02) & 0.93 (0.18) & 0.93 (0.19) & 0.90 (0.22) \\
         $g_2$=AMOS & liver & 0.95 (0.08) & 0.96 (0.08) & 0.98 (0.05) & 0.93 (0.14) \\
         & pancreas & 0.84 (0.26) & 0.85 (0.15) & 0.87 (0.20) & 0.78 (0.26) \\

        \hline
    \end{tabular}
\end{table}

\begin{table}[H]
\centering
\caption{Training dataset sizes by group ($g_1$ and $g_2$) and dataset from which they were sampled.}\label{experiments}
\begin{tabular}{|P{4cm}|P{3cm}|P{3cm}|}
\hline
\textbf{Groups} &  \textbf{AMOS} & \textbf{TotalSegmentator} \\
\hline
$g_1$=Female, $g_2$=Male & 380 & 150 \\ 
\hline
$g_1$=Under 50, $g_2$=Over 70 & 160 & 80 \\ 
\hline
$g_1$=TS, $g_2$=AMOS & 160 & 160 \\
\hline
\end{tabular}
\end{table}

%% file: Paper-0009.bbl
\begin{thebibliography}{10}
\providecommand{\url}[1]{\texttt{#1}}
\providecommand{\urlprefix}{URL }
\providecommand{\doi}[1]{https://doi.org/#1}

\bibitem{antonelli2022medical}
Antonelli, M., Reinke, A., Bakas, S., Farahani, K., Kopp-Schneider, A., Landman, B.A., Litjens, G., Menze, B., Ronneberger, O., Summers, R.M., et~al.: The medical segmentation decathlon. Nature communications  \textbf{13}(1), ~4128 (2022)

\bibitem{castro2020causality}
Castro, D.C., Walker, I., Glocker, B.: Causality matters in medical imaging. Nature Communications  \textbf{11}(1), ~3673 (2020)

\bibitem{Chen2022}
Chen, C., et~al.: Enhancing mr image segmentation with realistic adversarial data augmentation. Medical Image Analysis  \textbf{82} (2022)

\bibitem{chouker2004estimation}
Chouker, A., Martignoni, A., Dugas, M., Eisenmenger, W., Schauer, R., Kaufmann, I., Schelling, G., L{\"o}he, F., Jauch, K.W., Peter, K., et~al.: Estimation of liver size for liver transplantation: the impact of age and gender. Liver transplantation  \textbf{10}(5),  678--685 (2004)

\bibitem{gava2011gender}
Gava, A., Freitas, F., Meyrelles, S., Silva, I., Graceli, J.: Gender-dependent effects of aging on the kidney. Brazilian Journal of Medical and Biological Research  \textbf{44},  905--913 (2011)

\bibitem{gichoya2022ai}
Gichoya, J.W., Banerjee, I., Bhimireddy, A.R., Burns, J.L., Celi, L.A., Chen, L.C., Correa, R., Dullerud, N., Ghassemi, M., Huang, S.C., et~al.: Ai recognition of patient race in medical imaging: a modelling study. The Lancet Digital Health  \textbf{4}(6),  e406--e414 (2022)

\bibitem{guan2021domain}
Guan, H., Liu, M.: {Domain adaptation for medical image analysis: A survey}. IEEE Transactions on Biomedical Engineering  (2021)

\bibitem{hatamizadeh2022unetr}
Hatamizadeh, A., Tang, Y., Nath, V., Yang, D., Myronenko, A., Landman, B., Roth, H.R., Xu, D.: Unetr: Transformers for 3d medical image segmentation. In: Proceedings of the IEEE/CVF Winter Conference on Applications of Computer Vision. pp. 574--584 (2022)

\bibitem{hesamian2019deep}
Hesamian, M.H., Jia, W., He, X., Kennedy, P.: Deep learning techniques for medical image segmentation: achievements and challenges. Journal of digital imaging  \textbf{32},  582--596 (2019)

\bibitem{ioannou2022study}
Ioannou, S., Chockler, H., Hammers, A., King, A.P., Initiative, A.D.N.: A study of demographic bias in cnn-based brain mr segmentation. In: International Workshop on Machine Learning in Clinical Neuroimaging. pp. 13--22. Springer (2022)

\bibitem{Isensee2021}
Isensee, F., Jaeger, P.F., Kohl, S.A.A., Petersen, J., Maier-Hein, K.H.: nnu-net: a self-configuring method for deep learning-based biomedical image segmentation. Nature Methods  \textbf{18}(2),  203--211 (Feb 2021)

\bibitem{isensee2019nnu}
Isensee, F., Petersen, J., Kohl, S.A., J{\"a}ger, P.F., Maier-Hein, K.H.: nnu-net: Breaking the spell on successful medical image segmentation. arXiv preprint arXiv:1904.08128  \textbf{1}(1-8), ~2 (2019)

\bibitem{ji2022amos}
Ji, Y., Bai, H., Ge, C., Yang, J., Zhu, Y., Zhang, R., Li, Z., Zhanng, L., Ma, W., Wan, X., et~al.: Amos: A large-scale abdominal multi-organ benchmark for versatile medical image segmentation. Advances in Neural Information Processing Systems  \textbf{35},  36722--36732 (2022)

\bibitem{kipp2019normal}
Kipp, J.P., Olesen, S.S., Mark, E.B., Frederiksen, L.C., Drewes, A.M., Fr{\o}kj{\ae}r, J.B.: Normal pancreatic volume in adults is influenced by visceral fat, vertebral body width and age. Abdominal Radiology  \textbf{44},  958--966 (2019)

\bibitem{koh2021wilds}
Koh, P.W., Sagawa, S., Marklund, H., Xie, S.M., Zhang, M., Balsubramani, A., Hu, W., Yasunaga, M., Phillips, R.L., Gao, I., et~al.: Wilds: A benchmark of in-the-wild distribution shifts. In: International Conference on Machine Learning. pp. 5637--5664. PMLR (2021)

\bibitem{kreel1973changes}
Kreel, L., Sandin, B.: Changes in pancreatic morphology associated with aging. Gut  \textbf{14}(12),  962--970 (1973)

\bibitem{larrazabal2020gender}
Larrazabal, A.J., Nieto, N., Peterson, V., Milone, D.H., Ferrante, E.: Gender imbalance in medical imaging datasets produces biased classifiers for computer-aided diagnosis. Proceedings of the National Academy of Sciences  \textbf{117}(23),  12592--12594 (2020)

\bibitem{lee2022systematic}
Lee, T., Puyol-Ant{\'o}n, E., Ruijsink, B., Shi, M., King, A.P.: A systematic study of race and sex bias in cnn-based cardiac mr segmentation. In: International Workshop on Statistical Atlases and Computational Models of the Heart. pp. 233--244. Springer (2022)

\bibitem{Li2023well}
Li, W., Yuille, A., Zhou, Z.: How well do supervised models transfer to 3d image segmentation? In: International Conference on Learning Representations (2023)

\bibitem{ma2021cutting}
Ma, J.: Cutting-edge 3d medical image segmentation methods in 2020: Are happy families all alike? arXiv preprint arXiv:2101.00232  (2021)

\bibitem{marcos2015liver}
Marcos, R., Correia-Gomes, C., Miranda, H., Carneiro, F.: Liver gender dimorphism: insights from quantitative morphology. Histology and Histopathology  \textbf{30}(12),  1431--1437 (2015)

\bibitem{milletari2016v}
Milletari, F., Navab, N., Ahmadi, S.A.: V-net: Fully convolutional neural networks for volumetric medical image segmentation. In: 2016 fourth international conference on 3D vision (3DV). pp. 565--571. IEEE (2016)

\bibitem{Petersen2016}
Petersen, S.E., et~al.: {Reference ranges for cardiac structure and function using cardiovascular magnetic resonance (CMR) in Caucasians from the UK Biobank population cohort}. Journal of Cardiovascular Magnetic Resonance  \textbf{19}(1) (2016)

\bibitem{sabolic2007gender}
Saboli{\'c}, I., Asif, A.R., Budach, W.E., Wanke, C., Bahn, A., Burckhardt, G.: Gender differences in kidney function. Pfl{\"u}gers Archiv-European Journal of Physiology  \textbf{455},  397--429 (2007)

\bibitem{wang2023bias}
Wang, R., Chaudhari, P., Davatzikos, C.: Bias in machine learning models can be significantly mitigated by careful training: Evidence from neuroimaging studies. Proceedings of the National Academy of Sciences  \textbf{120}(6),  e2211613120 (2023)

\bibitem{wasserthal2022totalsegmentator}
Wasserthal, J., Breit, H.C., Meyer, M.T., Pradella, M., Hinck, D., Sauter, A.W., Heye, T., Boll, D., Cyriac, J., Yang, S., et~al.: Totalsegmentator: robust segmentation of 104 anatomical structures in ct images. arXiv preprint arXiv:2208.05868  (2022)

\bibitem{yang2023change}
Yang, Y., Zhang, H., Katabi, D., Ghassemi, M.: Change is hard: A closer look at subpopulation shift. arXiv preprint arXiv:2302.12254  (2023)

\bibitem{zhou2022domain}
Zhou, K., Liu, Z., Qiao, Y., Xiang, T., Loy, C.C.: {Domain generalization: A survey}. IEEE Transactions on Pattern Analysis and Machine Intelligence  (2022)

\bibitem{zhou2023multi}
Zhou, Y., Lee, H.H., Tang, Y., Yu, X., Yang, Q., Bao, S., Spraggins, J.M., Huo, Y., Landman, B.A.: Multi-contrast computed tomography atlas of healthy pancreas. arXiv preprint arXiv:2306.01853  (2023)

\end{thebibliography}
